\documentclass[twocolumn,epjc3]{svjour3}

\RequirePackage[T1]{fontenc}

\smartqed  

\RequirePackage{graphicx}
\RequirePackage{mathptmx}      
\RequirePackage{flushend}
\RequirePackage[numbers,sort&compress]{natbib}
\RequirePackage[dvipdfmx, bookmarksnumbered, pdfstartview=FitH,colorlinks,citecolor=blue,urlcolor=blue,linkcolor=blue]{hyperref}
\RequirePackage{amsmath}
\usepackage{amsmath}
\RequirePackage{overpic,graphicx}
\RequirePackage{dcolumn}
\RequirePackage{bm}
\RequirePackage{rotating}
\RequirePackage{subfigure}
\RequirePackage{color}
\RequirePackage{lineno}
\usepackage{hyperref}
\RequirePackage{multirow}
\usepackage{subcaption}
\usepackage{mathptmx}
\usepackage{enumerate}
\usepackage{tabularx}
\usepackage{changepage}
\DeclareMathAlphabet{\mathcal}{OMS}{cmsy}{m}{n}
\DeclareSymbolFont{largesymbols}{OMX}{cmex}{m}{n}

\journalname{Eur. Phys. J. C}
\begin{document}
\begin{sloppypar}



\title{Prospects of $|V_{us}|$ and axial vector form factors in $\Lambda\to pe^{-}{\bar\nu}_{e}$ decay at STCF}


\author{Junxian Zhou\thanksref{addr1, addr2}
        \and
        Shun Wang\thanksref{addr3, e1}
        \and
        Tao Luo\thanksref{addr1, addr2, e2}
        \and
        Xiaorong Zhou\thanksref{addr4, addr5, e3}
}

\thankstext{e1}{email: {wangshun@swust.edu.cn}
(corresponding author)}
\thankstext{e2}{email: {luot@fudan.edu.cn}
(corresponding author)}
\thankstext{e3}{email: {zxrong@ustc.edu.cn}
(corresponding author)}

\institute{Fudan University, Shanghai 200443, People's Republic of China\label{addr1}
          \and
Key Laboratory of Nuclear Physics on Ion-Beam Application (MOE) and Institute of Modern Physics, Fudan University, Shanghai 200443, People's Republic of China\label{addr2}
         \and
School of Mathematics and Physics, Southwest University of Science and Technology, Mianyang 621010, China
\label{addr3}
         \and
University of Science and Technology of China, Hefei 230026, People's Republic of China\label{addr4}
         \and
State Key Laboratory of Particle Detection and Electronics, Hefei 230026, People's Republic of China\label{addr5}
}

\date{Received: date / Accepted: date}

\maketitle

\begin{abstract}
We report a feasibility study of the semileptonic decay $\Lambda\to pe^{-} {\bar\nu}_{e}$ by using a fast simulation software package at STCF. With an anticipated integrated luminosity of 3.4 trillion $J/\psi$ per year at a center-of-mass energy at 3.097 GeV, the statistical sensitivity of the branching fraction is determined to be 0.15\%. The statistical sensitivities of form factors $g_{av}$ and  $g_w$ are determined to be 0.4\% and 2.15\%, respectively. Combining this result with $g_1(0)$ from Lattice QCD, we can obtain the projected sensitivity of $|V_{us}|$, to be 0.9\%, which is comparable to the precision obtained through meson decay measurements. The precise measurement to be obtained at STCF will provide a rigorous test of Standard Model.
\end{abstract}

\section{\boldmath{Introduction}}
\label{Intro}
The proposed Super Tau-Charm Facility (STCF)~\cite{2024STCF} in China is a symmetric electron-positron collider that will provide $e^{+}e^{-}$ annihilation at center-of-mass (c.m.) energies $\sqrt{s}$ ranging from 2.0 to 7.0 GeV. Its peak luminosity is expected to be $0.5\times10^{35}\mathrm{~cm}^{-2}\mathrm{s}^{-1}$ at $\sqrt{s}=$ 4.0 GeV and it will accumulate an integrated luminosity of more than 1~ab$^{-1}$ per year. As expected, we can obtain approximately 3.4 trillion $J/\psi$ events at a c.m. energy of 3.097 GeV with a one-year data collection, corresponding to $10^9$ $\Lambda\bar\Lambda$ pairs. This will enable researchers to study semileptonic and hadronic decays of $\Lambda$ with unprecedented precision.

Currently, the $|V_{us}|=0.2243\pm0.0008$ adopted by the Particle Data Group (PDG) is extracted from the kaon decay \cite{PhysRevD.110.030001}. Combining this $|V_{us}|$ result and the independently measured $|V_{ud}|$ and $|V_{ub}|$, the equation can be obtained $|V_{ud}|^2+|V_{us}|^2+|V_{ub}|^2=0.99848\pm0.00070$. There is a 2.3$\sigma$ tension with unitarity, leading to poor consistency of the Standard Model global fit. Moreover, with the result extracted from the $\tau$ decay \cite{PhysRevD.107.052008}, the equation can be obtained $|V_{ud}|^2+|V_{us}|^2+|V_{ub}|^2=0.99687\pm0.00087$, indicating a 3.6$\sigma$ tension with CKM matrix unitarity. Besides, the semileptonic decay $\Lambda\to pe^{-} {\bar\nu}_{e}$ also provides a window to determine the value of $|V_{us}|$. 
If the most precise measurement of $|V_{us}|=0.2250\pm0.0027$ from the hyperon decay is used  \cite{PhysRevLett.92.251803}, we can get $|V_{ud}|^2+|V_{us}|^2+|V_{ub}|^2=0.99879\pm0.00136$, thus in good agreement with unitarity. However, the result of the semileptonic decay experiment is very outdated and scarce and shows a high statistical uncertainty compared to the results of the kaon decay and $\tau$ decay \cite{PhysRevD.107.052008,PhysRevLett.92.251803,PDG}. Therefore, precise measurements of $|V_{us}|$ through semileptonic decays are imperative for rigorously testing the unitarity of the CKM matrix.

For the semileptonic hyperon decays $B_1\to B_2+\ell+\overline{\nu}_l$, using $M_1$ and $M_2$ to denote the masses of $B_1$ and $B_2$, the total decay rate for the electronic mode without radiative correction in the standard model, neglecting electron mass and setting $g_2(0)=0$ \cite{PhysRevLett.92.251803,citation-key}, is given by 
\\
\begin{equation}
\label{eq:Vus}
\frac{1}{\Gamma_{SM}}\frac{\mathrm{d}\Gamma_{SM}}{\mathrm{d}q^2}
=G_F^2\frac{\tau_\Lambda}{\hbar}\mathcal{K}|V_{us}\cdot f_1|^2\left[\mathcal{F}_0(q^2)+\delta\mathcal{F}_1(q^2)+\delta^2\mathcal{F}_2(q^2)\right],
\end{equation}
\\
where $\delta=\frac{(m_\Lambda-m_p)}{m_\Lambda}$, with $m_\Lambda$ and $m_p$ denoting the masses of the $\Lambda$ and proton, respectively. The weak decay constant $G_F$ and lifetime of $\Lambda$, $\tau_\Lambda$ can be obtained from PDG with high precision. The functions $\mathcal{F}_{k=0,1,2}$ depend on form factor ratios $g_{av}$, $g_{w}$, and $g_{av2}$, which represent the axial vector, weak magnetism and weak electricity couplings at zero momentum transfer $q^2$, respectively. The definition of $g_{av}$,$g_{w}$ and $g_{av2}$ are \cite{2009Helicity,PhysRevD.16.2165},
\\
\begin{equation}
g_{av}=\frac{g_1(0)}{f_1(0)},g_{w}=\frac{f_2(0)}{f_1(0)},g_{av2}=\frac{g_2(0)}{f_1(0)}.
\end{equation}
\\
$f_1(0)$, $g_2(0)$, and $f_2(0)$ are the vector, axial-vector, and weak-magnetism form factors at zero momentum transfer, respectively.
As the equation depends on $|V_{us}|$, if we can get the ${\Gamma_{SM}}$, $g_{av}$, $g_{w}$ from experiments and $g_1(0)$ from lattice QCD input, we can obtain the value of $|V_{us}|$.

Moreover, when radiative corrections are taken into account, the equation can be obtained \cite{PhysRevD.55.5702},
\\
\begin{equation}
\Gamma_{\mathbf{SM}}^c=\Gamma_{\mathbf{SM}}\times(1+Cr),
\end{equation}
\\
where $Cr$ is the radiative correction factor, calculated to be $Cr$ =$(1.6 \pm 0.5)\%$ for $\Lambda\to pe^{-} {\bar\nu}_{e}$ according to Ref.~\cite{PhysRevD.55.5702}.

In this research, we present a feasibility analysis of $\Lambda\to pe^{-} {\bar\nu}_{e}$and estimate the sensitivity of the branching fraction and $g_{w}, g_{av}$ at STCF. Since $g_{2}=0$ is assumed to be zero, $g_{av2}$ is not considered in this work. In our analysis, $\Lambda$ is from $e^+e^-\to J/\psi $, $J/\psi \to \Lambda\overline{\Lambda}$ at $\sqrt{s}$= 3.097 GeV. According to the conceptual design report, STCF is prospected to collect 3.4 trillion $J/\psi$ at $\sqrt{s}$=3.097 GeV \cite{2024STCF}. This yield is approximately 100 times greater than BESIII collaboration in terms of order of magnitude, which helps get results with high precision.

This paper is organized as follows. In Sect.~\ref{MC simu}, the detector concept for STCF has been introduced as well as our Monte Carlo (MC) samples used for this study. In Sect.~\ref{Ana}, the analysis of $\Lambda\to pe^{-} {\bar\nu}_{e}$ and the prospect of the branching fraction $\mathcal{B}_{\Lambda\to pe^{-} {\bar\nu}_{e}}$ is described. The calculation method for $g_{av}$ and $g_{w}$ along with their sensitivities, is detailed in Sect.~\ref{Form}. Optimization of the systemic uncertainty and response of the detector is elaborated in Sect.~\ref{Detector}. Finally, result and discussion of $|V_{us}|$ are shown in Sect.~\ref{Results}.

\section{\boldmath{MC simulation and STCF detector}}
\label{MC simu}
The STCF detector under development is a general-purpose detector for $e^+e^-$ collider. It consists of a tracking system composed of inner and outer trackers, a particle identification (PID) system, an electromagnetic calorimeter with good position resolution for photons or electrons, a muon detector that provides good $\mu/\pi$ separation. A detailed conceptual design for each sub-detector can be found in Ref.~\cite{2024STCF}.

Currently, the STCF detector and offline software system are in the research and development \cite{Li_2021}. Consequently, STCF has developed a fast simulation software for physics analysis , which takes the most common event generators as input to perform a realistic simulation. It takes into account the effects of charged particle tracking efficiency and momentum resolution, PID efficiency, and kinematic fits. The scaling factor of detector an be adjusted according to the performance limitations of the STCF detector and these configurations can be easily interfaced. Our target decay process $\Lambda\to pe^{-} {\bar\nu}_{e}$ can also serve as a benchmark process for the optimization of detector response, tracking efficiency, and $e/\pi$ separation capability.

A pseudo-data sample, corresponding to 1 billion $J/\psi$ events at $\sqrt{s}$ = 3.097 GeV was generated, which includes all open decay channels. The production of the $J/\psi$ resonance is simulated by the MC event generator KKMC \cite{PhysRevD.63.113009}, with effects of initial state radiation(ISR) and final state radiation(FSR) considered. In this pseudo-data sample, the decay modes with known branching fractions are included.

For our signal, we generate 4.5 million mDIY signal events using the formalism from Ref.~\cite{PhysRevD.108.016011}. As for the dominant background, $J/\psi\to\Lambda\bar{\Lambda},\Lambda\to p\pi^-,\bar{\Lambda}\to\bar{p}\pi^+$, we also use the mDIY generator according to Ref.~\cite{PhysRevD.108.016011}. All the parameter values we used to generate these exclusive MC are summarized in Table 1. The $\alpha_\psi$ governs the $\Lambda$ angular distribution and sin($\Delta\Phi$) is proportional to the hyperon polarization. $\alpha_\Lambda$ describes the transition from the $\Lambda$ to the proton \cite{Aebischer:2023mbz}. The maximum likelihood function is defined by these parameters. These values are all taken from the latest and most precise experimental measurements \cite{PhysRevLett.129.131801,PhysRevD.41.780}, while the value of $g_{w}$ is cited from the Cabibbo theory since the measurement now is not reliable
 \cite{PhysRevLett.129.131801,Cabibbo2003SEMILEPTONIC}. In this analysis, the $g_{av2}$ is assumed to be zero.


\begin{table}[!htbp]
\caption{The parameter values used to generate the signal MC \cite{PhysRevLett.129.131801,PhysRevD.41.780,Cabibbo2003SEMILEPTONIC}}

\label{tab:1}       
\begin{center}
\begin{tabular}{ccccccc}

\hline
$\alpha_\mathrm{\psi}$ & $\Delta\Phi$ & $\alpha_\mathrm{\Lambda}$ & $g_{av}^\mathrm{\Lambda}$ & $g_{w}^\mathrm{\Lambda}$  \\
\hline
0.4748 & 0.7521 & 0.7519 & 0.719 & 1.066 \\
\hline

\end{tabular}
\end{center}
\end{table}

The process of particles through the detector in this analysis is simulated by the fast simulation software \cite{Shi_2021}.

\section{\boldmath{Analysis of $\Lambda\to pe^{-} {\bar\nu}_{e}$}}
\label{Ana}
A double-tag technique is employed to precisely measure the absolute branching fraction of signal process $\Lambda\to pe^{-} {\bar\nu}_{e}$. It means that when a $\bar{\Lambda}$ is reconstructed, the presence of a ${\Lambda}$ is guaranteed. The $\bar{\Lambda}$ events are referred to as single-tag (ST) events. In the recoil system of the ST events, we can select the semileptonic decays of $\Lambda\to pe^{-} {\bar\nu}_{e}$(called double-tag (DT) events).
The ST and DT yields observed from data are given by 
\\
\begin{equation}
N_{\mathrm{ST}}=2N_{\Lambda\bar{\Lambda}}\mathcal{B}_{\mathrm{ST}}\epsilon_{\mathrm{ST}},    
\end{equation}
\\
and
\\
\begin{equation}
N_{\mathrm{DT}}=2N_{\Lambda\bar{\Lambda}}\mathcal{B}_{\mathrm{ST}}\mathcal{B}(\Lambda\to pe^{-} {\bar\nu}_{e})\epsilon_{\mathrm{DT},\Lambda\to pe^{-} {\bar\nu}_{e}},    
\end{equation}
where $N_{\Lambda\bar{\Lambda}}$ is the number of collected $\Lambda\bar{\Lambda}$ pairs ;$\mathcal{B}_{\mathrm{ST}}$ is the branching fraction of the single-tag mode. $\mathcal{B}(\Lambda\to pe^{-} {\bar\nu}_{e})$ is the branching fraction of the double-tag mode. $\epsilon_{ST}$ and $\epsilon_{DT}$ are the efficiency of reconstructing the ST mode(called the ST efficiency) and the efficiency of simultaneously collecting the ST mode and DT mode (called the DT efficiency). Based on these two equations, the equation for the absolute branching fraction of $\Lambda\to pe^{-} {\bar\nu}_{e}$ can be derived as 
\\
\begin{equation}
\label{eq:DTmethod}
 \mathcal{B}(\Lambda\to pe^{-}\bar\nu_e)=\frac{N_{\mathrm{DT}}/\epsilon_{\mathrm{DT},\Lambda\to pe^{-}\bar\nu_e}}{N_{\mathrm{ST}}/\epsilon_{\mathrm{ST}}},   
\end{equation}
\\
with the DT method described above, we can get the absolute branching fraction. Throughout this letter, the charge conjugation can be implied if needed with similar method \cite{PhysRevLett.127.121802}.

All candidate charged tracks are selected if they pass the vertex in fast simulation. The $\bar{\Lambda}$ candidates are reconstructed by $\bar{p}$ and $\pi^+$ with the vertex-constrained fit to a common point. The $\chi^2$ primary vertex fit is required to be smaller than 100, which is a common and loose requirement to suppress the potential non-$\bar{\Lambda}$ background events. Moreover, we also require the track pairs have oppositely charges and polar angle cut $|\cos(\theta)|<0.93$, where $\theta$ is the polar angle with respect to the $z$ axis, which is the axis of the MDC. Since $\Lambda$ and $\bar{\Lambda}$ are long-life particles, the decay length divided by its uncertainty obtained from the second vertex fit is required to be $L/\sigma>2$. At least one $\bar{\Lambda}$ hyperon is required successfully reconstructed after performing primary vertex and second vertex fit. To distinguish the $\bar{\Lambda}$ from combinatorial background, we define the beam-energy constrained mass of tagged $\bar{p}$$\pi^+$ system as follows.
\\
\begin{equation}
    M_{\mathrm{BC}}=\sqrt{E_{\mathrm{beam}}^2-|\vec{p}_{\bar{p}\pi^+}|^2},
\end{equation}
\\
where $E_{beam}$ is the beam energy and $|p_{\bar{p}\pi^+}|$ is the magnitude of the momentum of the daughter particle $\bar{p}\pi^+$ system after vertex fit. When we consider the equation in the $J/\psi$ frame, the magnitude of $\Lambda$ and $\bar{\Lambda}$ should be equal. The conclusion can be deduced as follows:
\\
\begin{equation}
    E_{\mathrm{beam}}=E_{\Lambda}=E_{\bar{\Lambda}}=\frac{E_{\mathrm{cms}}}{2}=\frac{3.097}{2}GeV.
\end{equation}
\\
Since there may be several different combinations for the ST mode, it is important to select the best combination from candidates. So we define $\Delta E$ to identify the tagged $\bar{\Lambda}$ candidates. First, we calculate the energy difference 
\\
\begin{equation}
    \Delta E=E_{\bar{\Lambda}}-E_{\mathrm{beam}},
\end{equation}
\\
where $E_{\bar{\Lambda}}$ is the reconstructed energy of a tagged $\bar{\Lambda}$ and $E_{\rm beam}$ is the beam energy. 
\begin{figure}[h!]
    \centering
    \includegraphics[width=0.9\linewidth]{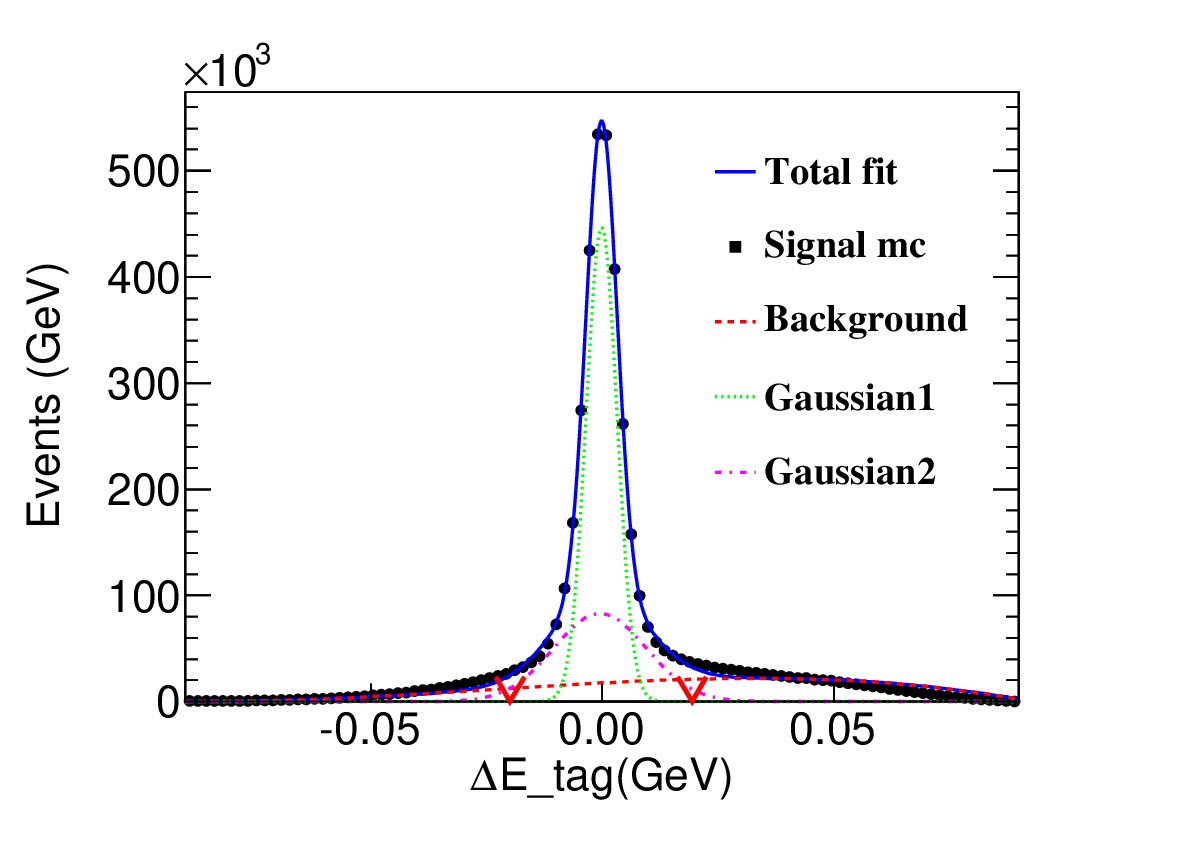}
     \begin{flushleft}
    \caption{The distribution of $\Delta E$ in signal MC.The black dots with error bar represent pseudo-data.The blue line is a total fitting result. The red dashed line is the background from wrong combination. The green dashed line and deep pink dashed line represent the Gaussian function to fit respectively.}
    \label{}
     \end{flushleft}
\end{figure}
If there are multiple candidates for ST mode, one with the minimum $|\Delta E|$ is retained for further analysis. $\Delta E$ cut($-0.019<\Delta E<0.021\mathrm{~GeV}/c^2$) can be shown in Fig.~\ref{1}. The mismatch between the observed fits and the MC is due to background contamination from partial miscombinations. Considered the resolution of the detector, the peak of $E_{\bar{\Lambda}}$ should be around zero in the $\Delta E$ distribution. To improve the signal purity, requirements of $\Delta E$ are applied, which is approximately $\pm3\sigma_{\Delta E}$ around zero. 
\begin{figure}[htp]
    \centering
    \includegraphics[width=0.8\linewidth]{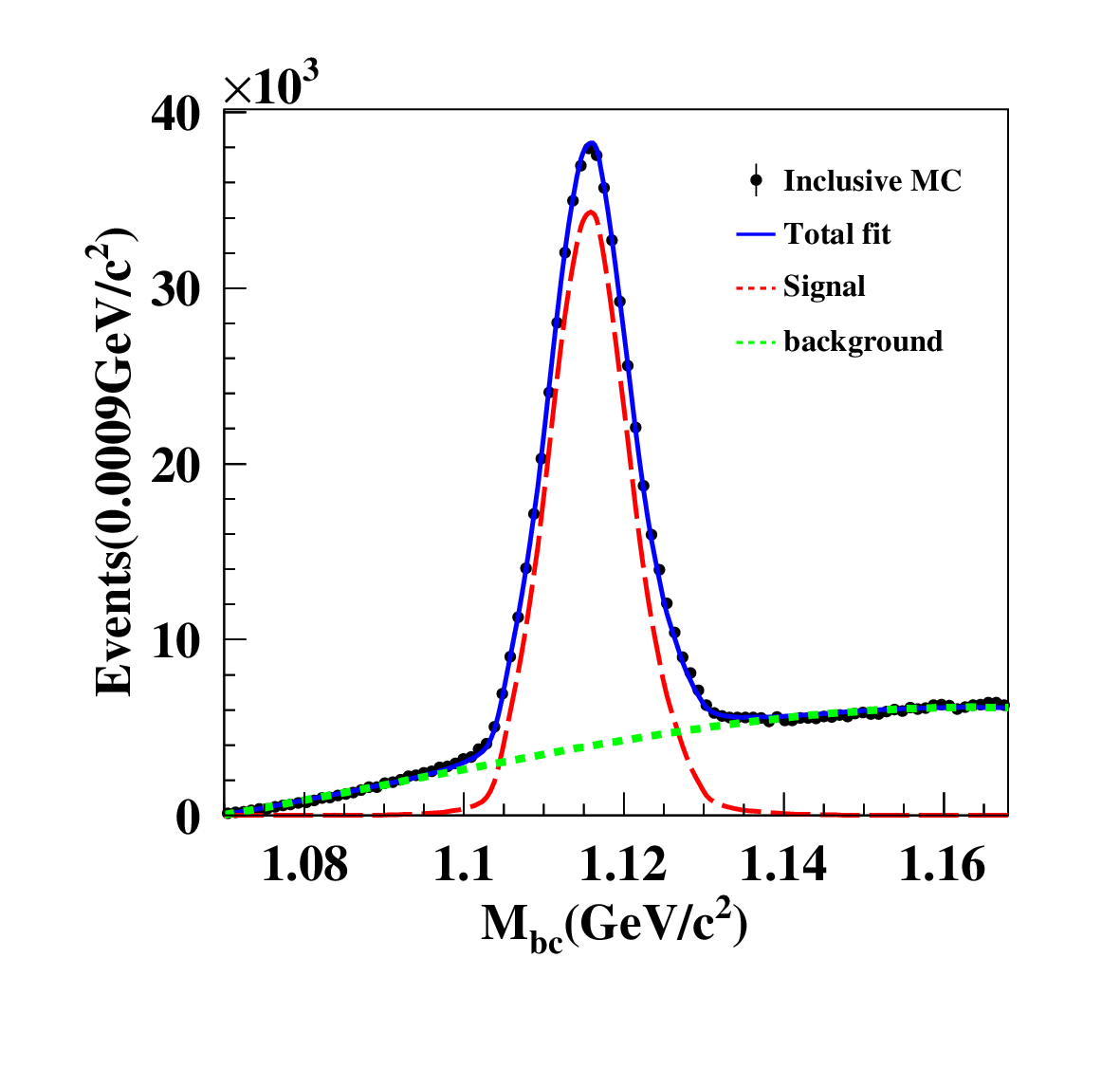}
     \begin{flushleft}
    \caption{Fit to the $M_{BC}$ distribution of pseudo-data .The black dots with error bar represent total fitting result. The red dashed line is signal shape modeled with the MC-simulated shape convolved with a Gaussian function. The green dashed line is background shape modeled with a third-order Chebyshev function.}
    \label{fig:mmmmm}
     \end{flushleft}
\end{figure}
After these selection criteria, we use $M_{BC}$ to get $N_{tag}$. Figure.~\ref{fig:mmmmm} shows the $M_{BC}$ distribution, where the signal is modeled using a MC-determined signal shape convolved with a Gaussian function.To select the signal process with a high purity, a mass window is applied on $M_{BC}$ within $\pm3\sigma_{M_{\mathrm{BC}}}$, where $\sigma_{M_{\mathrm{BC}}}$ is determined by signal MC shape. It can be defined as $(1.110, 1.142)\mathrm{~GeV}/c^2$. We can get the resolution of $M_{BC}$ by fitting with a double-Gaussian function. According to the result of fits, we can extract the $N_{tag}$ and $\epsilon_{tag}$.
Totally, we obtain $N_{tag}=479810\pm800$ entries. Through our input pseudo-data, we can get numbers of pairs of $\Lambda\overline{\Lambda}$. Combining these results, $\epsilon_{tag}=(37.93\pm0.06)\%$ is obtained.

After selecting the ST mode, we can select the DT mode. We require exactly one candidate $\bar{\Lambda}$ at ST mode.Hence, based on the 2 good tracks at ST mode, we require another 2 good tracks at DT side to reconstruct ${\Lambda}$ with the criteria for additional good charged tracks the same as those used in the ST selection. Since $\bar{\nu_e}$ cannot be detected by detectors at STCF, the ${\Lambda}$ is reconstructed through $p$ and $e^-$. To obtain good tracks with sufficient quality, $\chi^2$ vertex selection, decay length selection, and PID selection criteria are applied. The PID likelihoods calculated by ionization energy loss, time-of-flight and electromagnetic calorimeter information satisfy $\mathcal{L}_e>0.001$ and $\frac{\mathcal{L}_e}{\mathcal{L}_e+\mathcal{L}_\pi+\mathcal{L}_\mu}>0.8$, where $\mathcal{L}_e$,$\mathcal{L}_e$ and $\mathcal{L}_e$ are likelihoods calculated based on fast simulation \cite{Shi_2021}.For additional background reduction, optimization functions implemented in the fast simulation software were employed, which will be detained introduced in Section.~\ref{Detector}. As the neutrino is not detected, we choose the kinematic quantity of the neutrino as fit quantity \cite{PhysRevLett.127.121802}.

To obtain the information about the missing neutrino, the kinematic quantity is defined as 
\\
\begin{equation}
U_{\mathrm{miss}}\equiv E_{\mathrm{miss}}-c|\vec{p}_{\mathrm{miss}}|
\end{equation}
\\
where $E_{miss}$ and $p_{miss}$ are the total energy and the momentum of all missing particles in the event, respectively. $E_{\rm miss}$ is calculated by 
\begin{equation}
E_{\mathrm{miss}}=E_{\mathrm{beam}}-E_p-E_{e^-},
\end{equation}
\\
where $E_{\mathrm{beam}}$ is the beam energy, $E_p$ and $E_{e^-}$ are the measured energies of $p$ and $e^-$, respectively. We can use constrained $\Lambda$ momentum to calculate $p_{miss}$ and $\vec{p}_{\Lambda}$
\\
\begin{equation}
p_{\mathrm{miss}}=|\vec{p}_\Lambda-\vec{p}_p-\vec{p}_{e^-}|, 
\\
\vec{p}_{\Lambda}=-\frac{\vec{p}_{\mathrm{tag}}}{|\vec{p}_{\mathrm{tag}}|}\sqrt{E_{\mathrm{beam}}^{2}-m_{\Lambda}^{2}},
\end{equation}
\\
where $\vec{p}_{\Lambda}$ is the momentum of the tagged $\bar{\Lambda}$ hyperon, and $m_{\Lambda}$ is the nominal $\Lambda$ mass. The fit to the distribution of $U_{miss}$ is shown in Fig.~\ref{fig:enter-label}.

\begin{figure}[htp]
    \centering
    \includegraphics[width=0.85\linewidth]{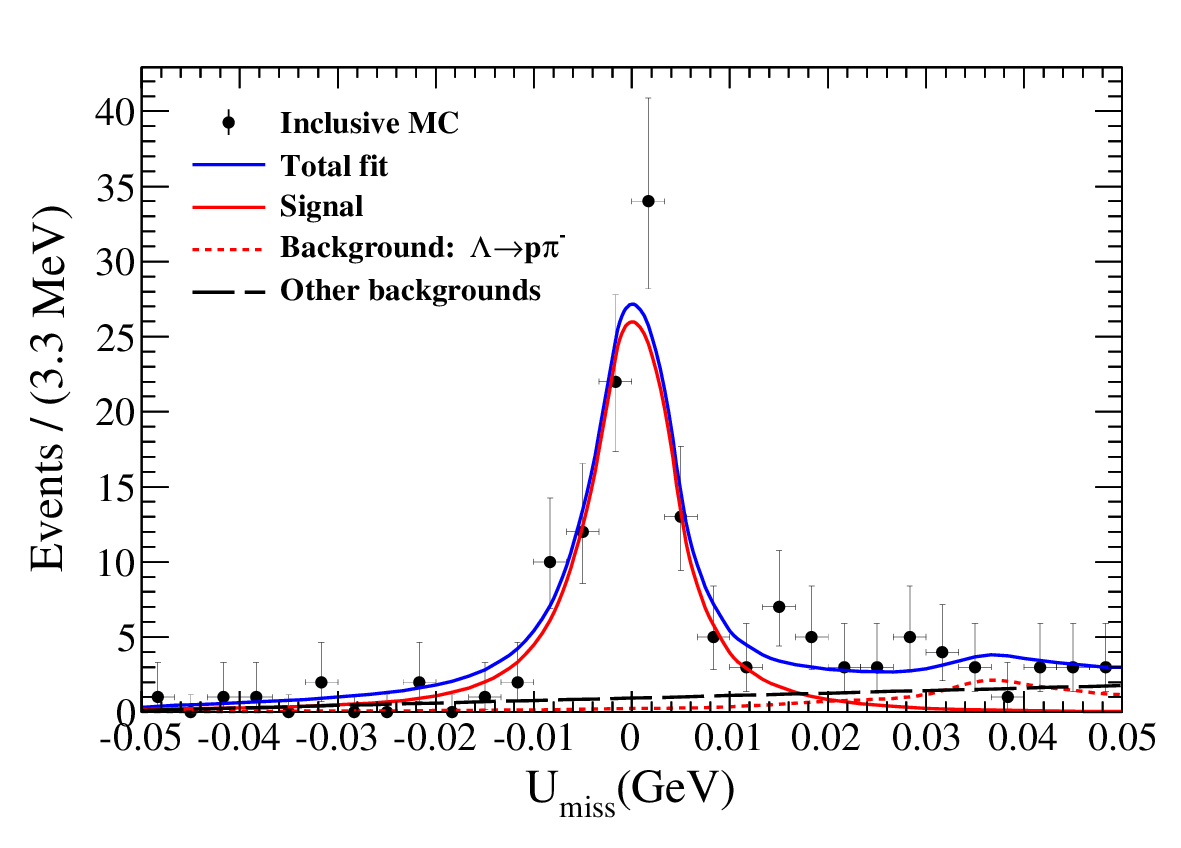}
    \caption{Fit to the $U_{\rm miss}$ distribution of pseudo-data .The black dots with error bars represent pseudo-data.The blue line is a total fitting result. The red line is signal shape modeled with the MC-simulated shape convolved with a Gaussian function. The red dashed line is the dominate peek background $\Lambda\to p\pi^{-}$ The black dashed line is background shape modeled with a first-order Chebyshev function.}
    \label{fig:enter-label}
\end{figure}

Through the fit to $U_{\rm miss}$, we can obtain $N_{sig}=104\pm 12$. With all the same selection criteria for our signal MC, $\epsilon_{sig}$ is calculated to be $14.13\%$. So we can calculate the BF of $\Lambda\to pe^{-} {\bar\nu}_{e}$ by Eq.~\eqref{eq:DTmethod}, to be $\mathcal{B}(\Lambda\rightarrow pe^{-}\bar{\nu}_{e})=(6.12\pm0.61)\times10^{-4}$. The uncertainties are just statistical uncertainties from $N_{sig}$. The relative uncertainty is approximately $9.99\%$. The calculated BF is consistent with the input value. We can prospect the relative statistical sensitivity for the BF of $\Lambda\to p e^{-} {\bar\nu}_{e}$ at STCF with 3.4 trillion $J/\psi$ as it scales with $1/\sqrt{\mathcal{L}}$, to be about $0.17\%$ \cite{2022Prospects}.

\begin{figure}[htp]
    \centering
    \includegraphics[width=0.8\linewidth]{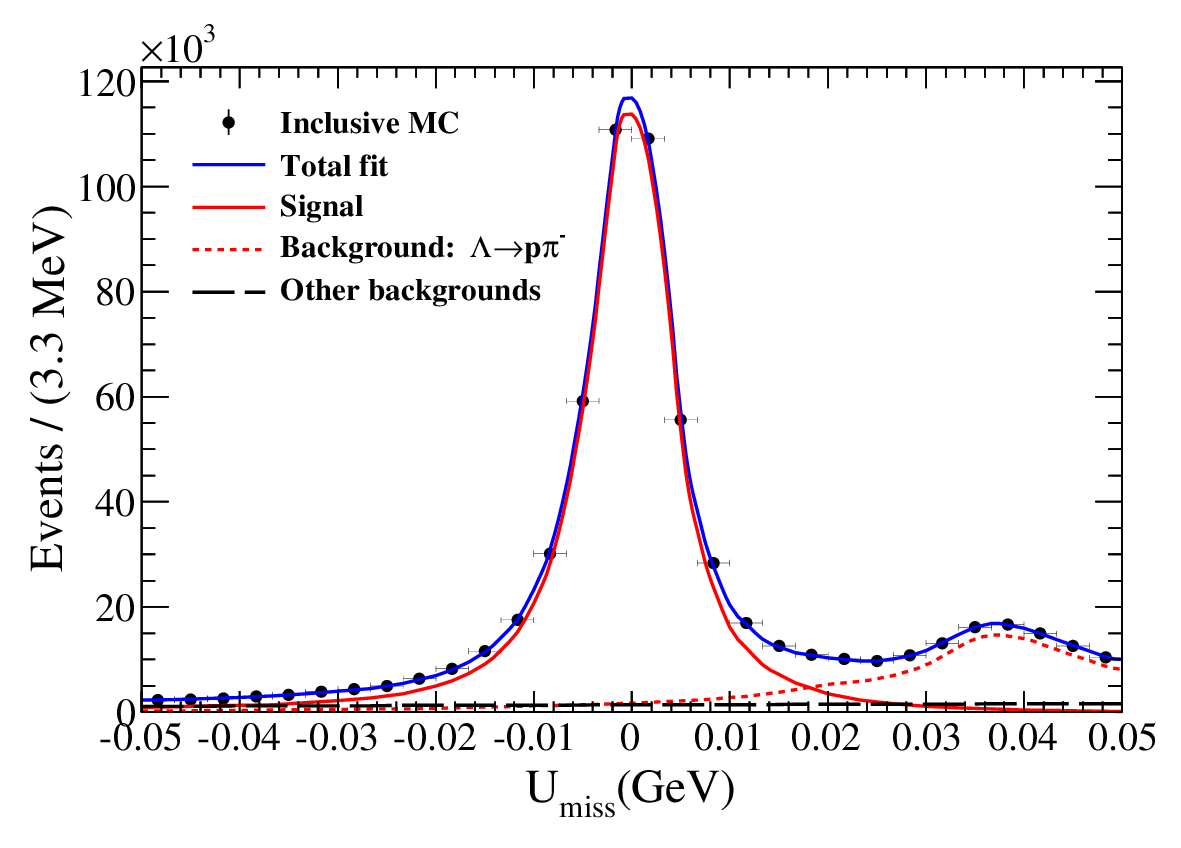}
    \caption{Fit to the $U_{\rm miss}$ distribution of the new pseudo-data .The black dots with error bar represent pseudo-data.The blue line is a total fitting result. The red line is signal shape modeled with the MC-simulated shape convolved with a Gaussian function. The red dashed line is the dominate peek background $\Lambda\to p\pi^{-}$ The black dashed line is background shape modeled with a first-order Chebyshev function.}
    \label{fig:enter-label}
\end{figure}

To confirm our prospect, we use signal MC and the background MC after bootstrap method to make a new pseudo-data sample, which can be used to estimate the fitting result when collecting 3.4 trillion $J/\psi$. Considered only the statistical uncertainties, through the fit shown in Fig.~\ref{fig:enter-label}, the relative uncertainty of the branching fraction is estimated to be $0.167\%$, which is consistent with our prospect through the proportion between BF and $1/\sqrt{\mathcal{L}}$. In further analysis , this relative uncertainty will be used to calculate the $|V_{us}|$. The systematic uncertainty is not included since the construction of the detectors is not complete. It will be discussed in detail in Section \ref{Results} \cite{2022Prospects,Fan:2021mwp,Li:2021ala}.

\section{\boldmath{Form factor measurement and calculation of $|V_{us}|$}}
\label{Form}

Since one billion pseudo-data cannot fully represent the situation of statistical sensitivity at 3.4 trillion $J/\psi$ at STCF \cite{2024STCF}, we use the new pseudo-data which we use to confirm the relative uncertainty in the end of Sect.~\ref{Ana} to estimate the statistical sensitivity of form factor. To extract more pure signal events, slightly modified selection criteria are applied.

\begin{figure}[htp]
    \centering
    \includegraphics[width=1.05\linewidth]{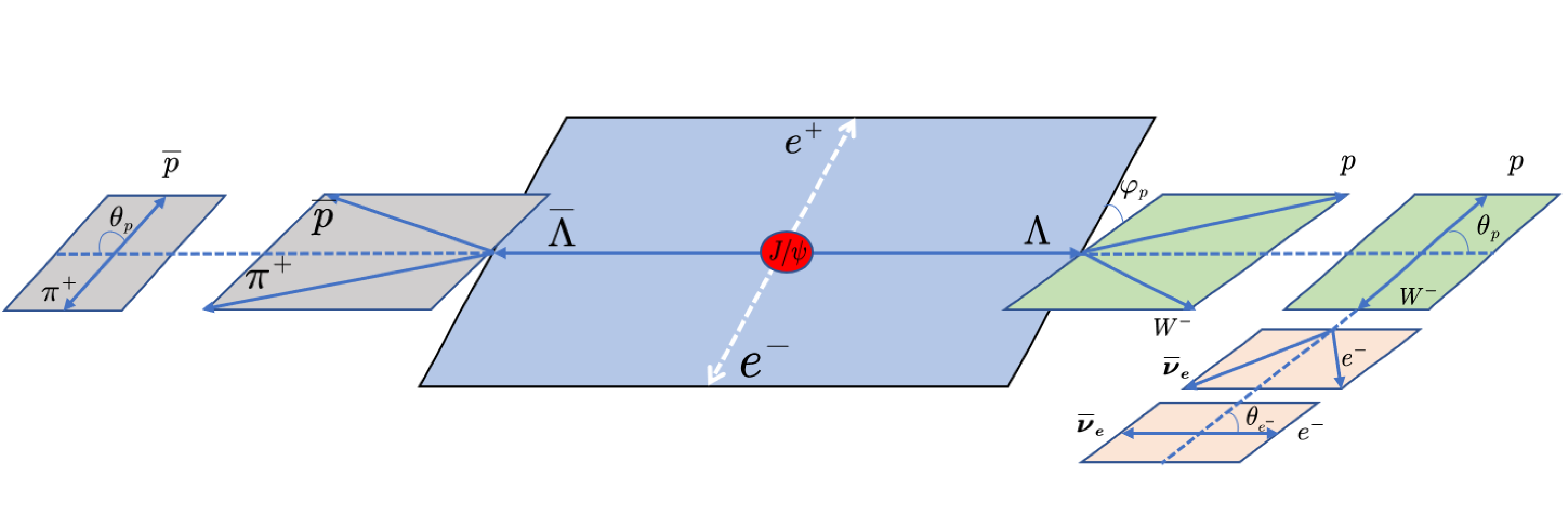}
        \caption{Definition of the helicity angles. In our fit, these helicity angels are converted to the corresponding centroidal frame for analysis. }
    \label{fig:helicity}
\end{figure}

The Maximum Log-Likelihood method is used to measure the form factor. Firstly the function of the kinematic variables can be defined as $\xi=(\theta,\theta_p,\varphi_p,\theta_e,\theta_{\bar{p}},\varphi_{\bar{p}},q^2)$, including six helicity angles and the four-momentum transfer between $\Lambda$ and p. The helicity angles are obtained by boosting particles into their respective helicity frames as illustrated in Fig.~\ref{fig:helicity}.
\\
\\

At the level of our prospects, charge parity is conserved symmetry, so we can obtain $\alpha_-=-\alpha_+,g_{a\nu}^-=-g_{a\nu}^+,g_w^-=g_w^+$. Based on this symmetry, the six-dimensional angular distribution is determined by six global parameters, which is written as $\Omega=(\alpha_\psi,\Delta\Phi,g_{av},g_w,g_{av2},\alpha_+)$. Since we assume $g_2(0)=0$ , these parameters can be reduced to five parameters as listed in Table 1.
\\
\begin{equation}
    \mathcal{L}=\prod_{i=1}^N\mathrm{Prob}(\xi_i)=\prod_{i=1}^N\frac{W(\xi_i;\Omega)}{C},\quad C=\frac{1}{N_{mc}}\sum_{j=1}^{N_{mc}}\frac{W(\xi_j;\Omega)^{mc}}{W(\xi_j;\Omega_0)^{mc}},
\end{equation}
\\
Prob(${\xi}_i$) is the probability of the event $i$ characterized by the measurement ${\xi}_i$; W is the differential cross-section. $\Omega_{0}$ is the set of parameters we used to generate the mDIY MC, with its input parameters detailed in Table 1. $\Omega$ is our target fitting variables. To subtract the contributions from inclusive backgrounds and $p\pi$ backgrounds, the maximum likelihood function is written as 
\\
\begin{equation}
-\ln\mathcal{L}_{sig}=-\ln\mathcal{L}_{data}+\ln\mathcal{L}_{bkg-p\pi}+\ln\mathcal{L}_{bkg-other}.   
\end{equation}
\\
In addition to the selection criteria for calculating $\mathcal{B}(\Lambda\rightarrow pe^{-}\bar{\nu}_{e})$, we require $\begin{vmatrix}U_{\mathrm{miss}}\end{vmatrix}<0.02\mathrm{~GeV}$ to purer signal.This requirement suppresses almost all backgrounds except $p\pi$ background. Since we need to prospect the relative uncertainty at STCF with 3.4 trillion $J/\psi$. We generate a pseudo-data sample using the same method as in Section 3. The elements in the sample are mixed according to the ratio of events after cut. We can use mDIY MC of different elements to perform precise calculations respectively and determine the minimum sum of these likelihood functions finally.
In our fit , we fix these parameters $\alpha_\mathrm{\psi}$,$\Delta\Phi$,$\alpha_\mathrm{\Lambda}$ and make our target form factors $g_{av}^\mathrm{\Lambda}$,$g_{w}^\mathrm{\Lambda}$ float.
\begin{figure*}[htp!]
\centering  
\subfigure[result of $g_{av}$]{
\label{Fig.sub.1}
\includegraphics[width=6.95cm,height = 5.8cm]{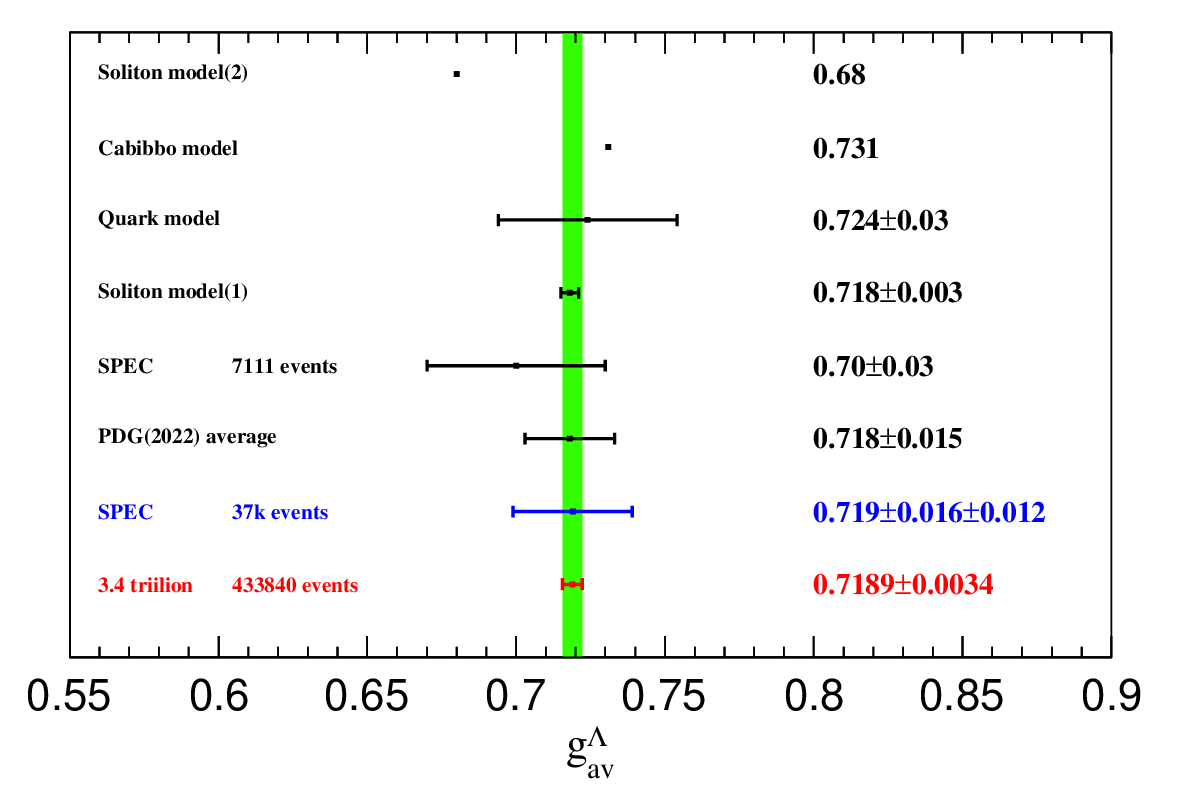}}\subfigure[result of $g_{w}$]{
\label{Fig.sub.2}
\includegraphics[width=6.95cm,height = 5.8cm]{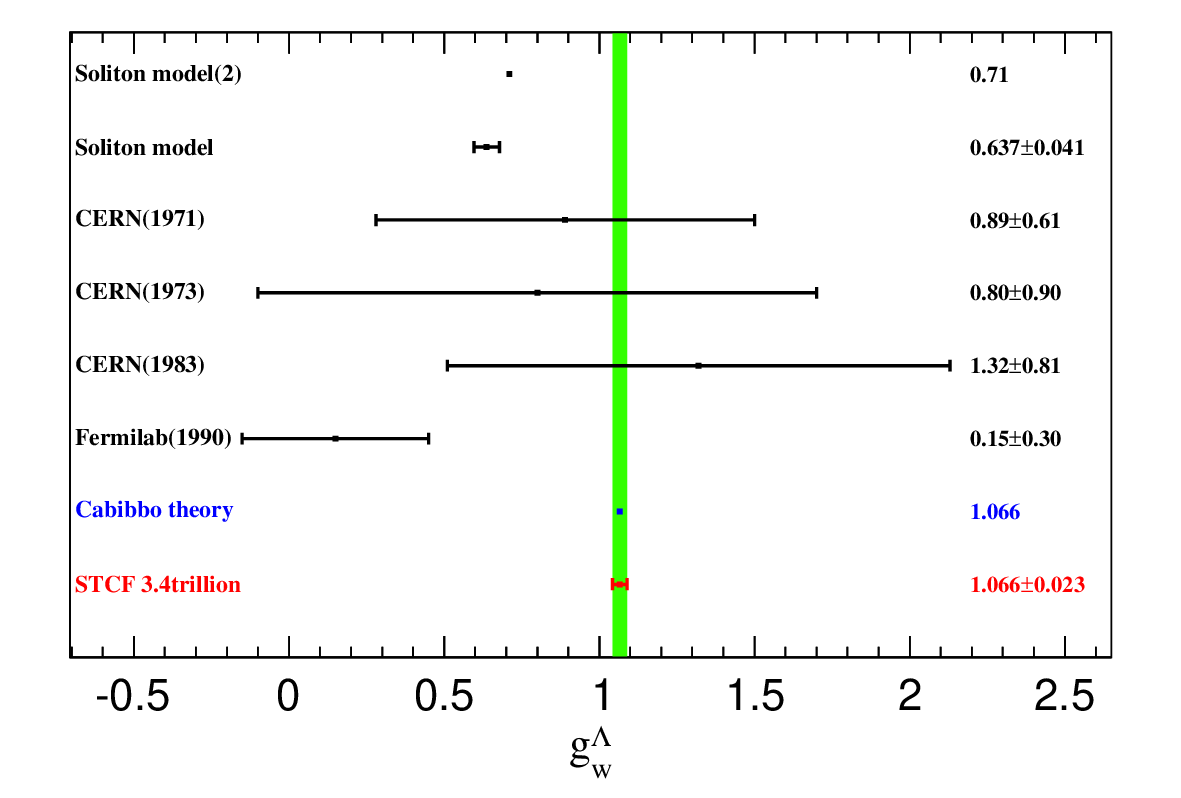}}
\caption{the prospect of $g_{av}$ and $g_w$. These red parts represents our analysis results, and the blue part represents our input values. Black parts represent other measurements or theoretical calculation results.}
\label{gavgw}
\end{figure*}

Through these fittings, we can obtain our prospect of $g_{av}^\mathrm{\Lambda}$, $g_{w}^\mathrm{\Lambda}$ to be $g_{av}=0.7189\pm0.0034$ and $g_{av}=1.066\pm0.023$, the uncertainties are only statistical. The comparison of current measurements and theoretical calculation of  $g_{av}^\mathrm{\Lambda}$,$g_{w}^\mathrm{\Lambda}$ with our prospect are shown in Fig.~\ref{gavgw} \cite{Faessler:2008ix,Yang:2015era,Ledwig:2008ku,WISE1981123,B}.

The relative statistical uncertainty of the $g_{av}$ and $g_{w}$ are estimated to be 
0.5\% and 2.1\% respectively. Compared with current measurements, the statistical uncertainty is reduced by approximately tenfold. These results may also be able to exclude some theoretical outcomes such as soliton model at high precision if STCF can be constructed completely.

To extract $f_1(0)$ through $g_{av}$, a LQCD input $g_1(0)$ is required. In this analysis, $g_1(0)$ is cited according to Ref.~\cite{Bacchio:2025auj},
 \\
 \begin{equation}
 g_1(0)=-0.8263\pm0.0070.   
 \end{equation}
 \\
This latest result takes into account the effects of SU(3) symmetry breaking. As we obtain the expected value of the $\mathcal{B}(\Lambda\rightarrow pe^{-}\bar{\nu}_{e})$ and $g_{av}$,$g_{w}$ in this analysis, the LQCD input $g_1(0)$, the world average values of $G_F$,$m_{\Lambda}$,$m_{p}$,$\tau_{\Lambda}$, the value of $|V_{us}|$ can be obtained according to Eq.~\ref{eq:Vus},
\\
\begin{equation}
    |V_{us}|=0.2335\pm 0.0021.  
\end{equation}
\\
The source of uncertainty for this value can be divided in four parts: $\mathcal{B}(\Lambda\rightarrow pe^{-}\bar{\nu}_{e})$ obtained in this analysis, $g_{av}$ and $g_{w}$ obtained in this analysis, $g_1(0)$ from LQCD, and parameters from PDG. The values input and uncertainties are summarized in Table~\ref{tab:666}. The total uncertainty can be obtained through
\\
\begin{equation}
    \Delta_{sum}=\sqrt{\sum_{i=1}^{4}\Delta_{i}^2},
\end{equation}
\\
where $\Delta_{i}$ is divided into 4 parts displayed in the fourth column of Table~\ref{tab:666}.
The uncertainties of values from our analysis are only statistical. The total relative uncertainty is about 8.99\%. Significantly, the LQCD input constitutes the dominant source of uncertainty in this result, accounting for approximately 90.48\%. More precise theoretical results are anticipated in the future.
\begin{table}[!htbp]
\caption{The input value and their contribution to the uncertainty of the final result.$\delta_{i}$ means relative uncertainty of this part.}

\label{tab:666}       
\begin{center}
\begin{tabular}{p{1.0cm}p{2.8cm}p{1.8cm}p{1.3cm}}

\hline
Source & Input value & $\delta_{i}$(\%) & contribution to $\Delta_{V_{us}}$  \\
\hline
$\mathcal{B}_{DT}$ & $8.32\times10^{-4}$ & 0.17 & $0.0002_{\rm stat.}$  \\
\hline
$g_{av}$ & 0.7189 & 0.47 & \multirow{2}{*}{$0.0004_{\rm stat.}$}  \\
$g_{w}$  & 1.066  & 2.12 \\
\hline
$G_F$ &$1.1664\times10^{-5}$ GeV/$c^2$ & $5.14\times10^{-5}$ &\multirow{5}{*}{$0.0007$}\\
$m_{\Lambda}$ & 1.1157 GeV/$c^2$ & $5.38\times10^{-4}$\\
$m_{p}$ &0.9382 GeV/$c^2$ & $3.09\times10^{-8}$\\
$\tau_{\Lambda}$ &$2.6170\times10^{-10}$ s & $0.38$ \\
\hline
$g_1(0)$ & -0.8329 & 0.84& 0.0019\\
\hline
Sum& & &0.0021\\
\hline
\end{tabular}
\end{center}
\end{table}

\section{\boldmath{Detector response optimization and systematic uncertainty estimation}}
\label{Detector}
Since this analysis is based on the fast simulation for STCF \cite{Shi_2021}, 
a series of optimizations on detector responses have been performed in the results presented above, including the efficiency of charged tracks and PID efficiencies. In the fast simulation, by default, all parameters for each sub-detector performance are parameterized based on the BESIII performance \cite{BESIII:2009fln}, but can be adjusted flexibly by a scale factor according to expected performance of the STCF detector, or by implementing a special interface to model any performance described with an external histogram, an input curve, or a series of discrete data. With the help of the functions of fast simulation package, two kinds of detector response are studied below.

\begin{enumerate}[a.]
    \item \textit{Tracking efficiency} The tracking efficiency in fast simulation is characterized by two variables: transverse momentum $P_T$ and polar angle $\cos\theta$, which are correlated with the level of track bending and hit positions of tracks in the tracker system. For low-momentum tracks ($P_T < 0.2$ GeV/c), the reconstruction efficiency is low due to stronger electromagnetic multiple scattering, electric field leakage, energy loss. However, with different techniques in the tracking system design at STCF and track finding algorithm in development, the efficiency is expected to be improved for low-momentum tracks. In fast simulation, the detection efficiency is scaled with a factor ranging from 1.1 to 1.5 to estimate the optimization of tracking detector. For high-momentum tracks, the tracking efficiency within acceptance is over 99\%. Hence this variation has a little influence on the ST efficiency. Figure.~\ref{Fig.sub.1} illustrates the performance of tracking efficiency with the change of the charge scale factor. The red point is the default result, to be 1.0, which is used in Section.~\ref{Ana}. The larger scale factor is not used in our prospect since it also increases the dominated background $\Lambda \to p\pi^-$.
\end{enumerate}

\begin{enumerate}[b.]
    \item \textit{PID identification} At STCF, the $dE/dx$ from the tracking system is mainly used to separate $\pi/e$. The relation of the misidentification rate to the momentum/direction is estimated by geant4 simulation with the BESIII detector case. This relationship is inherited  in the fast simulation, which is used to estimate the misidentification rates for other momenta. In our analysis, misidentification from a pion to an electron when the momentum is less than 0.4 GeV / c forms the main peaking background mentioned above. As fast simulation provides the functions which allow us to vary the $\pi$/e misidentification rate for $\pi$/e, which is quantified by a factor R varying from -1 to 0.0001 in this article. The default misidentification rate R is to be -1.0. In Section.~\ref{Ana}, the PID cut at the condition of R=0.0001 is used to exclude the peaking background $\Lambda \to p\pi^-$. The test has been done to the function about $\pi/e$ identification as Fig~.\ref{Fig.sub.2} shows. When the factor R is varied to 0.0001, the $\pi$/e misidentification rate is scanned to approximately 0.07\%, which is consistent with the expected misidentification rate of STCF.
\\
\begin{figure*}[htp!]
\centering  
\subfigure[charge scale optimization]{
\label{Fig.sub.1}
\includegraphics[width=6.6cm,height = 4.7cm]{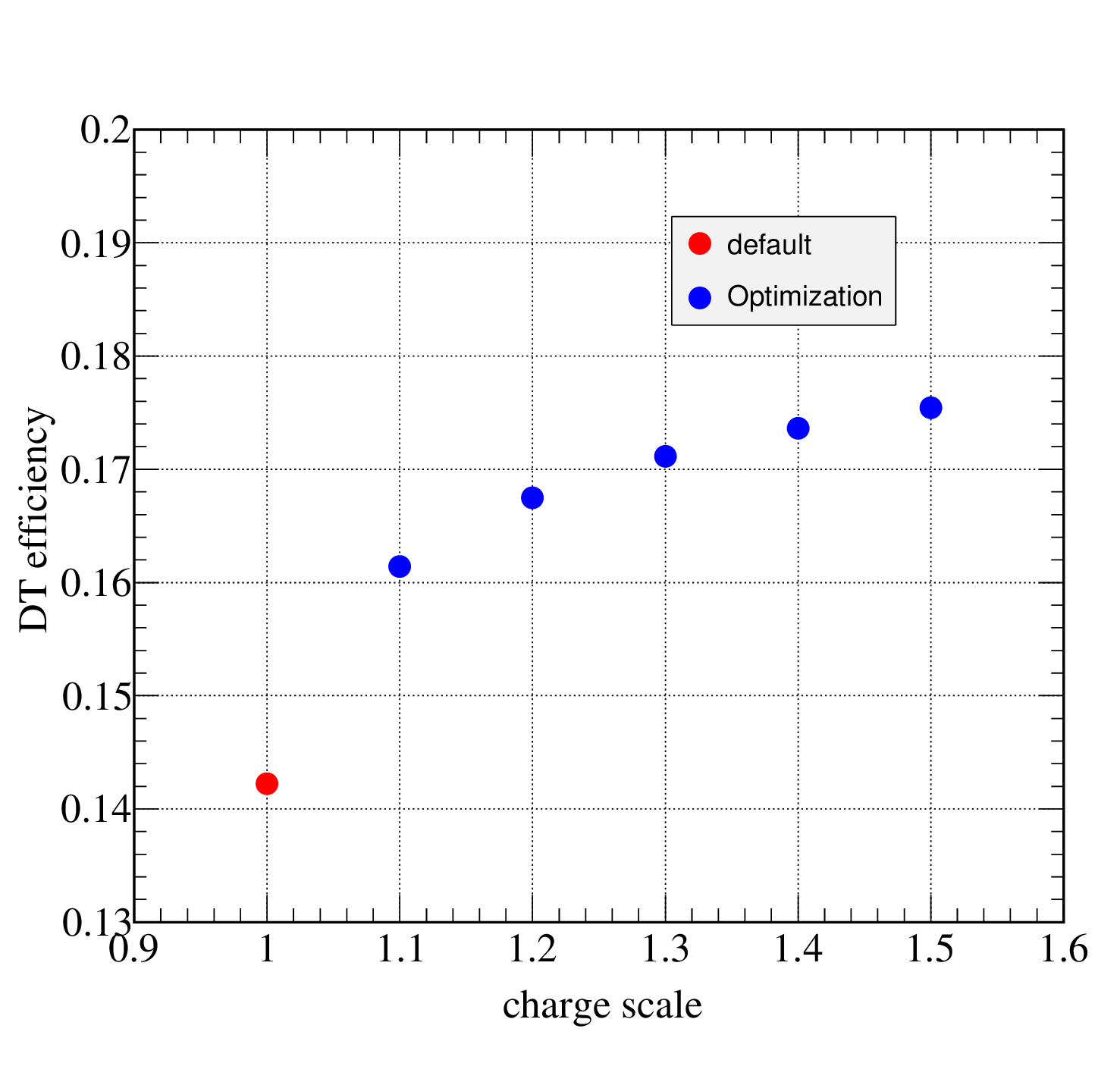}}\subfigure[PID identification optimization]{
\label{Fig.sub.2}
\includegraphics[width=6.6cm,height = 4.8cm]{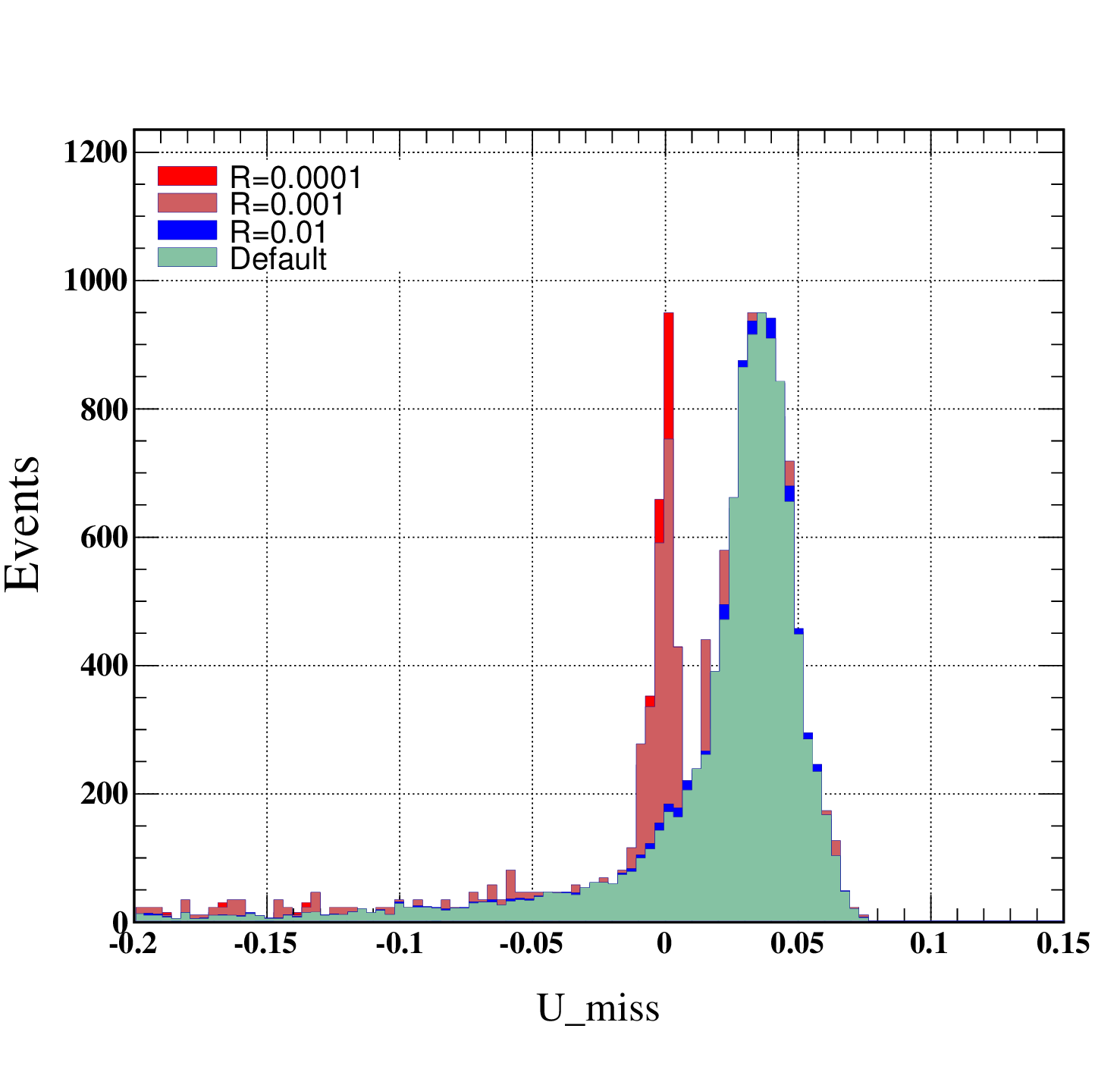}}
\caption{Detector response optimization through fast simulation.(a) describes the optimization of DT efficiency for charged tracks of reconstructed efficiency, the red star denotes the default resul.(b) describes the distribution of $\mathrm{U_{miss}}$ from inclusive MC with different R values. The red distribution is used in Sect~.\ref{Ana} Since the default distribution has low figure-of-merit.}
\label{1}
\end{figure*}
\\
\end{enumerate}

In  summary, these sets of optimization factors for different sub-detector responses are beneficial to get a more precise prospect of $|V_{us}|$ with higher DT efficiency and low misidentification. These fast simulation-based optimizations can also provide reference and optimization for the operational indicators of STCF in the specific detector design process.

Since a full systematic study requires both experimental data and MC, a precise estimation of systematic
uncertainty is not feasible until the construction of the detectors is completed. Therefore, a rough estimation is given by referring to similar measurements from the similar study at BESIII which is to be published \cite{PhysRevLett.127.121802,Shun}. The systematic uncertainties are classified into 2 cases named $\delta_{red}$ and $\delta_{inred}$.

The first one is related with the statistics of controlled samples, which are reducible with the expected luminosity. These systematic uncertainties can be normalized with the luminosity. In BESIII, the systematic uncertainty are studied mainly by the samples of $J/\psi \to \Lambda\bar{\Lambda} \to \bar{p}\pi^+p\pi^-$ through approximate $10^{10}$ $J/\psi$. Hence the systematic uncertainty can be estimated through scaling the control samples to $3.4*10^{12}$ $J/\psi$, whose sources are $N_{track}=4$, reconstruction through vertex fit, tracking of p, electron detection and kinematic fit.
The tracking and PID efficiencies of $e^+$ from the control sample can be used to correct the data/MC differences and estimate the corresponding systematic uncertainties. By scaling the control sample used at BESIII, the tracking of p and PID uncertainties of e are estimated to be 0.01\% and 0.08\%, respectively. Kinematic and vertex fit uncertainties are studied through the control sample  $J/\psi \to \Lambda\bar{\Lambda} \to \bar{p}\pi^+p\pi^-$ after the same selection criteria \cite{tong}. After scaling, these systematic uncertainties are estimated to be 0.01\%, 0.01\%. The uncertainty of $N_{\rm track}=4$ can be negligible after scaling.
 
The second one is related with the method of fit proceduce, which is predictably optimized through detector optimization in many aspects but cannot be estimated precisely at this time through the luminosity. Therefore these systematic uncertainties have been made a conservative estimate according to the study at BESIII. The uncertainty in fit to the $M_{\rm BC}$ of ST $\bar{\Lambda}$, is estimated by varying fit range, bin size, background shape, and signal shape for MC and data to be about 0.37\%. Similarly, the uncertainty in the fit to the $U_{\rm miss}$ of DT ${\Lambda}$ is estimated to be 0.8\% \cite{Shun}.

Finally, the systematic uncertainties of branching fraction are summarized in Table~\ref{tab:555}. The total systematic uncertainty can be roughly estimated as $\delta_{\mathrm{syst.}}=\sqrt{\delta_{\mathrm{red.}}^2+\delta_{\mathrm{irred.}}^2}$. As a result, the absolute branching fraction is estimated to be 
\\
\begin{equation}
    \mathcal{B}(\Lambda \to pe^-\bar{\nu}_e)=[8.32\pm0.01(\mathrm{stat.})\pm0.07(\mathrm{syst.})]\times10^{-4}.
\end{equation}

\begin{table}[!htp]
\caption{The relative systematic uncertainties in the measurement of the BF $\Lambda \to pe^-\bar{\nu}_e$.}
\label{tab:555}
\centering
\resizebox{1.0\linewidth}{!}{
\begin{tabular}{ccc}
\hline
Category & Source & relative uncertainty(\%) \\
\hline
\multirow{5}{*}{$\delta_{\rm red.}$} & proton tracking & 0.01 \\
 & e PID & 0.08 \\
 & Kinematic fit & 0.01 \\
 & $\Lambda$ reconstruction through vertex fit & 0.01 \\
 & $N_{\rm track}=4$ & negligible \\
\hline
\multirow{2}{*}{$\delta_{\rm irred.}$} & Fitting $M_{\rm bc}$ & 0.37 \\
 & Fitting $U_{\rm miss}$ & 0.80 \\
\hline
Sum & & 0.88 \\
\hline
\end{tabular}}
\end{table}

Notably, the dominated sources from fitting $M_{\rm bc}$ and $U_{\rm miss}$ will be improved in the future with the detector optimization \cite{2024STCF}.
The uncertainties of fitting method used in the measuring $g_{av}$ and $g_{w}$ are also considered conservatively according to the study at BESIII. After considering fixed parameters and background events, the relative uncertainties are estimated to be 0.57\% and 15.91\%, respectively.Therefore, the final results of $g_{av}$ and $g_{w}$ are estimated to be $g_{av}=0.7189\pm0.0034(\mathrm{stat.})\pm0.0041(\mathrm{syst.})$ and $g_{w}=1.066\pm0.023(\mathrm{stat.})\pm0.170(\mathrm{syst.})$. Here, the main sources come from the measurements of fixed parameters, which will be replaced with more precise measurements in the future.
Since the systematic uncertainties are rough estimations, these uncertainties will not be considered in the results given in the final discussion Sect~.\ref{Results}.

\section{\boldmath{Summary and discussion}}
\label{Results}
\begin{figure}[htp]

    \centering
    \includegraphics[width=1.0\linewidth]{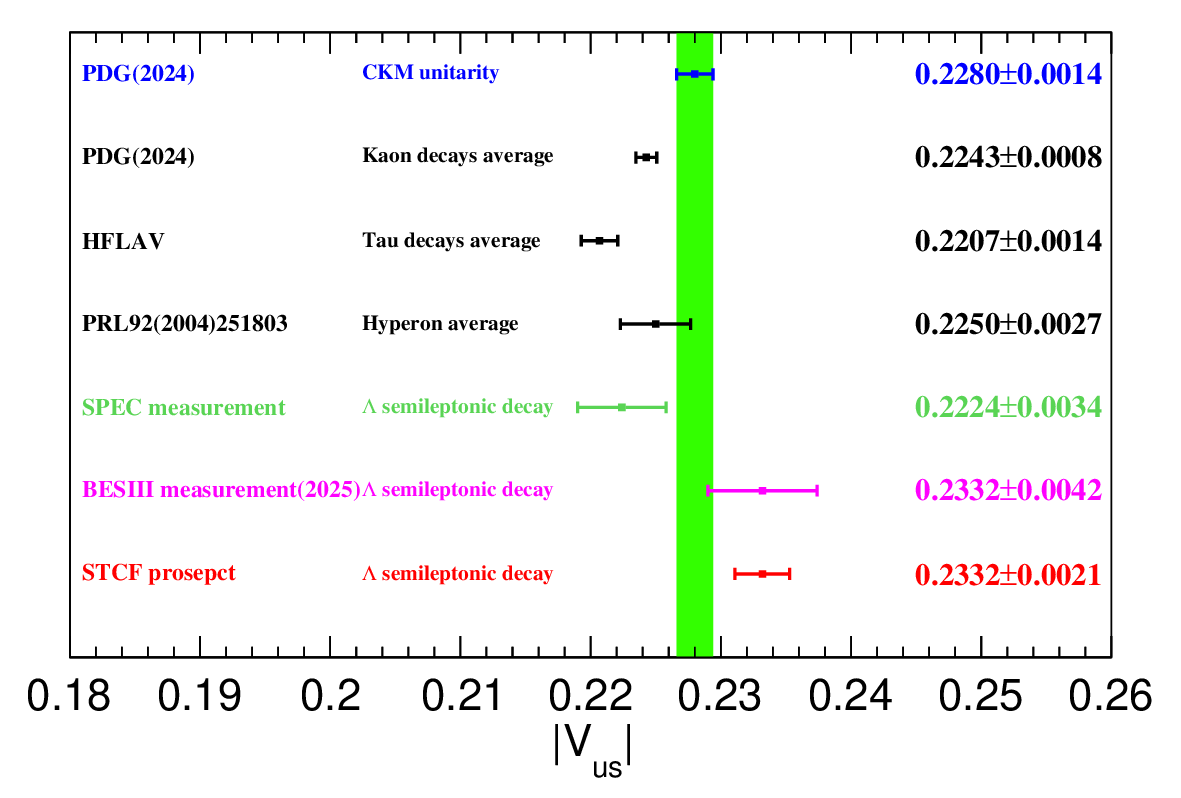}
    \caption{The comparison of STCF prospect with other measurements \cite{PhysRevD.107.052008,PhysRevD.110.030001,Cabibbo2003SEMILEPTONIC,PhysRevLett.92.251803,Shun}. The red part is the result with the assumption of the same center value.}
    \label{fig:Vus11}
\end{figure}

With these expected values mentioned in Section~\ref{Ana},\ref{Form} or cited from PDG \cite{PDG}, the predicted value of $|V_{us}|$ is expected to be $0.2335\pm 0.0021 $, where the statistical uncertainty is much smaller compared with statistical uncertainties of current measurement from hyperon decay. Currently, the most precise measurement of $|V_{us}|$ from $\Lambda\to pe^{-} {\bar\nu}_{e}$ is $0.2224\pm 0.0034 $, which shows a 1.4$\sigma$ deviation from the CKM unitarity \cite{PhysRevLett.92.251803,Cabibbo2003SEMILEPTONIC}.
Recently, the BESIII collaboration obtained a measurement using the same lattice QCD inputs as in this work, which is consistent with the CKM matrix values with comparatively large uncertainties \cite{Shun}. Since this article mainly focuses on uncertainties at STCF, the central value is assumed to align with the latest result. Leveraging high-luminosity environment at STCF, the final prospect of $|V_{us}|$ is to be $0.2332\pm 0.0021$. This prospect shows a 2.3$\sigma$ deviation from CKM unitarity, which can indicate the tension with the unitarity of the CKM matrix. In comparison, the uncertainty is suppressed to nearly half of its previous value. It is worth noting that the dominant source of systematic uncertainty in this analysis arises from the input LQCD values, accounting for approximately 90\%. Therefore, with future improvements in LQCD precision and operation of STCF detectors, it should be feasible to extract $|V_{us}|$ from hyperon decays with a precision comparable to that from kaon decays. If this tension persists in the future, it could point to evidence of new physics. The most recent experimental measurements, along with our theoretical predictions, are summarized in Fig.~\ref{fig:Vus11}.

Furthermore, the systematic uncertainties of $\mathcal{B}(\Lambda\rightarrow pe^{-}\bar{\nu}_{e})$, $g_{av}$, and $g_{w}$ are not included in this result since the detectors and software systems are not complete. However, the uncertainties of LQCD input and these parameters cited from PDG will also be reduced in the future with other more precise measurements or theories. Therefore, the precision of $|V_{us}|$ is expected to be more accurate in the future. In summary, these prospects suggest that hyperon decay experiments at STCF will provide stringent tests of CKM unitarity, potentially allowing for the search for new physics beyond Standard Model.
\\
\section*{\boldmath {Acknowledgments}}
This work is supported by the National Key R$\&$D Program of China under Contract No.2022YFA1602200 and the international partnership program of the Chinese Academy of Sciences Grant No.211134KYSB20200057. We thank the Hefei Comprehensive National Science Center for their strong support on the STCF key technology research project.


\bibliographystyle{spphys}
\bibliography{main}{}


\clearpage
\end{sloppypar}
\end{document}